\documentclass{aa}
\usepackage{psfig}

\newcommand{\nh}{N_{\rm H}}
\newcommand{\cmsq}{{\rm atoms\,cm}^{-2}}
\newcommand{\ergs}{{\rm erg\,s}^{-1}}
\newcommand{\ergcm}{{\rm erg\,cm}^{-2}}
\newcommand{\ergcms}{{\rm erg\,cm}^{-2}{\rm s}^{-1}}
\newcommand{\ltap}{\mathrel{\hbox{\rlap{\lower.55ex \hbox {$\sim$}}
                   \kern-.3em \raise.4ex \hbox{$<$}}}}
\newcommand{\gtap}{\mathrel{\hbox{\rlap{\lower.55ex \hbox {$\sim$}}
                   \kern-.3em \raise.4ex \hbox{$>$}}}}
\newcommand{\SAXI}{SAX\,J2224.9$+$5421}
\newcommand{\SAXII}{SAX\,J1828.5$-$1037}
\newcommand{\SAXIII}{SAX\,J1324.5$-$6313}

\begin{document}


\title{BeppoSAX Wide Field Cameras observations of six type\,I 
X-ray bursters}

\author{R.\ Cornelisse\inst{1,2} \and  F.\ Verbunt\inst{2} \and
	J.J.M.\ in 't Zand\inst{2,1} \and E.\ Kuulkers\inst{1,2} 
	\and J.\ Heise\inst{1} \and R.A. Remillard\inst{3}
	\and M. Cocchi\inst{4} \and L. Natalucci\inst{4} 
	\and A. Bazzano\inst{4} \and P. Ubertini\inst{4}}
\offprints{R.\ Cornelisse}
\mail{R.Cornelisse@sron.nl}

\institute{SRON National Institute for Space Research, Sorbonnelaan 2, 
              3584 CA Utrecht, The Netherlands  
         \and  Astronomical Institute, Utrecht University,
              P.O.Box 80000, 3508 TA Utrecht, The Netherlands
	\and Center for Space Research, Massachusetts Institute of
	     Technology, Cambridge, MA 02139-4307
	\and Istituto di Astrofisica Spaziale (CNR), Area Ricerca Roma
	      Tor Vergata, Via del Fosso del Cavaliere, I-00133, Roma, Italy}

\date{accepted May 15, 2002}   
 
\abstract{We have discovered three certain (SAX\,J1324.5$-$6313, 
2S\, 1711$-$339 and \SAXII) and two likely (SAX\,J1818.7$+$1424 and
\SAXI) new thermonuclear X-ray burst sources with the BeppoSAX Wide Field
Cameras, and observed a second burst ever from a sixth one (2S\,0918$-$549).
Four of them (excluding 2S\,1711$-$339 and 2S\,0918$-$549)
are newly detected X-ray sources from which we observed single bursts,
but no persistent emission. We observe the first 11 bursts 
ever from 2S\,1711$-$339; persistent flux was detected during the
first ten bursts, but not around the last burst. A single burst was
recently detected from 2S\,0918$-$549 by Jonker
et~al. (2001); we observe a second burst showing radius expansion,
from which a distance of 4.2\,kpc is derived.  According to theory,
bursts from very low flux levels should last $\gtap100$\,s. Such is
indeed the case for the last burst from 2S\,1711$-$339, the single
burst from \SAXII\ and the two bursts from 2S\,0918$-$549, but
not for the bursts from SAX\,J1324.5$-$6313, SAX\,J1818.7$+$1424
and \SAXI. The bursts from the latter sources all last
$\sim${20}\,s. We suggest that SAX\,J1324.5$-$6313,
SAX\,J1818.7$+$1424, \SAXII\ and \SAXI\ are members of the recently
proposed class of bursters with distinctively low persistent flux
levels, and show that the galactic distribution of this class is
compatible with that of the standard low-mass X-ray binaries.
}

\maketitle

\section{Introduction}

About 40\%\ of the low mass X-ray binaries in our Galaxy occasionally
show (so-called Type\,I) X-ray bursts, which are thermonuclear flashes
due to unstable helium and/or hydrogen burning of matter accreted on a
neutron star surface (for a review see e.g. Lewin et al. 1993).  A
typical burst shows a fast rise ($\simeq1$\,s) and exponential decay,
softening during the decay (interpreted as cooling of the neutron star
photosphere), and a spectrum which can be well described with
black-body radiation.  At the moment, about 70 X-ray bursters are
known, approximately 20 of which have been discovered with BeppoSAX
(e.g., in 't Zand 2001).

Most X-ray bursters are detected with persistent X-ray flux observed
before and after bursts. But some sources have been detected during
bursts only, with upper-limits on the persistent emission. These
limits vary widely, as they depend on the sensitivity of the
instrument used.  For example, the X-ray sources in the globular
clusters Terzan 1 and Terzan 5 were detected with the Hakucho
satellite during bursts only, but EXOSAT and ROSAT also detected the
persistent emission (Makishima et al.\ 1981, Warwick et al.\ 1988,
Verbunt et al.\ 1995).

The study of X-ray bursts serves various purposes. First, applying the
theory of X-ray bursts to observations provides information about the
neutron star (e.g., its radius when the distance is known) and about
its companion (e.g., whether the matter transferred from the companion
is hydrogen-rich or not).  Second, the bursts  unambiguously
decide which of the low-mass X-ray binaries contain a neutron star as
opposed to a black hole, and provide (an upper limit to) the distance
of the binary, from the condition that its luminosity should be less
than the Eddington luminosity.  Third, new low-mass X-ray binaries may
be discovered by the detection of bursts, in cases where the
persistent flux is too low.  The last two points help in forming a
more accurate view of the total number and distribution of low mass
X-ray binaries in our Galaxy, and of the fraction that contains a
neutron star.

In this paper we describe the observation with the BeppoSAX Wide Field
Cameras of  six type I burst sources. Four of these
are new X-ray sources with a persistent flux below the detection
threshold of the Wide Field Cameras of $\simeq10^{-10}$ $\ergcms$. 
The fifth source is a known
X-ray source, from which we detect bursts for the first time.  The
sixth source is also a previously known X-ray source, the first
burst of which was recently discovered by Jonker et al. (2001); we
describe a second burst from this source and use it to determine its
distance.  In Sect.\,2 we describe the observations and data
reduction; the results are described in Sect.\,3. Because of
their diversity, the sources are discussed in separate subsections,
each of which is accompanied by a sub-subsection in which comparison
with other observations are made. Finally, in Sect.\ 4 we discuss some
implications of our results for the theory of bursts (Sect. 4.1) and
for various sub-populations of the low-mass X-ray binaries with
neutron stars (Sect. 4.2).

\begin{table*}
\caption{Results of the BeppoSAX Wide Field Cameras observations.  For
each burst source the table gives the time of the burst, the position
with error $\delta$, the total exposure time on the source
between August 1996 and December 2001 $t_{\rm tot}$, the exposure time
$t_{\rm exp}$ of the pointing in which the burst was detected, and the
hydrogen absorption column $\nh$, persistent flux $F_{\rm pers}$
between 2-28 keV, and the distance $d$ derived from the burst peak
flux.  For the bursts the table gives the e-folding times $\tau$ for
the total flux, and for the hard and soft energies, where we choose
bands 2-x and x-28 keV such that both bands have similar
countrates. Spectral fits have been made for counts integrated over
$t_{\rm fit}$; we give the black body temperature $kT_{\rm bb}$,
radius $R$ at distance (limit) $d$, the average flux $F$ in two
bands, the bolometric peak flux $F_{\rm peak}$, and the total burst
fluence $E_{\rm b}$; and the temperature $kT_{\rm brems}$ and photon
index $\Gamma$ for bremsstrahlung and power law fits, respectively.
All fluxes are corrected for absorption.
\label{results}}  
\begin{tabular}{l@{\hspace{0.12cm}}c@{\hspace{0.23cm}}c@{\hspace{0.23cm}}c@{\hspace{0.23cm}}c@{\hspace{0.23cm}}c@{\hspace{0.23cm}}c@{\hspace{0.23cm}}}
\hline
& SAX  &SAX & SAX &SAX&2S&2S\\
& J1324.5$-$6313 & J1818.7+1424 & J1828.5-1037 & J2224.9$+$5421 
& 1711$-$339 & 0918$-$549\\ 
\hline
\multicolumn{6}{l}{\bf Source parameters}\\
Burst time (MJD) & 50672.151 & 50683.770 & 51988.863 & 51488.454 &b1-11, Table\,\ref{1711} & 51335.049\\
RA (J2000)  & $13^{\rm h} 24^{\rm m} 27^{\rm s}$ & $18^{\rm h} 18^{\rm m} 
44^{\rm s}$ & $18^{\rm h} 28^{\rm m} 33^{\rm s}$ & $22^{\rm h} 24^{\rm m} 
52^{\rm s}$ & $17^{\rm h} 14^{\rm m} 17^{\rm s}$ & $09^{\rm h} 20^{\rm m} 
37^{\rm s}$\\
Dec (J2000) & $-63^\circ 13\farcm4$ & $14^\circ 24\farcm2$ & $-10^\circ 
37\farcm8$  & $+54^\circ 21\farcm9$ & $-34^\circ3\farcm3$  & $-55^\circ 
13\farcm9$ \\
$\delta$ (99\% confidence.)       & $1\farcm8$& $2\farcm9$ & $2\farcm8$ & 
$3\farcm2$ & $1\farcm5$& $0\farcm7$\\
$l_{II},b_{II}$ & $306\fdg6, -0\fdg6$ & $42\fdg2,+13\fdg7$ &
$20\fdg9,0\fdg2$ & $102\fdg6,-2\fdg6$ & $352\fdg1, +2\fdg8$ & 
$275\fdg9, -3\fdg8$ \\
$t_{\rm tot}$ (day) & 58 & 31 & 25 & 65  & 66 & 62  \\
$t_{\rm exp}$ (ks) & 18.5 & 16.2 & 12.7 & 40.2 & 314 & 36 \\
$\nh$ ($10^{22}$ $\cmsq$)  & 1.5$^a$ & 0.1$^a$ & 1.9$^a$ & 0.5$^a$ 
& 1.5$^b$  & 0.24$^c$ \\
$F_{\rm pers}$ ($10^{-10}$ $~\ergcms$)& $<0.80$ ($3\sigma$)& $<1.7$ 
($3\sigma$) & $<1.9$ ($3\sigma$) & $<0.35$ ($3\sigma$) &$6.3\pm0.6$& 
$3.8\pm0.6$\\
$d$ (kpc)        & $<6.2$  & $<9.4$ & $<6.2$ & $<7.1$ &$<7.5$ & 4.2 \\ 
\hline
\multicolumn{5}{l}{\bf Burst parameters}\\
$\tau_{2-28 \rm keV}$ (s)             & $6.0\pm0.1$       & $4.5\pm0.1$ & 
 11.2$\pm$0.6 & 2.6$\pm$0.2 & $7.1\pm0.2$& $48.5\pm0.2$\\
x     & 8                 & 4           & 7 & 6 & 6          & 5\\
$\tau_{2-x\rm keV}$  (s)              & $9.7\pm0.4$       & $5.7\pm0.2$ & 
21.5$\pm$1.3 &  3.9$\pm$0.7 & $7.6\pm0.3$& $80.6\pm0.7$\\
$\tau_{x-28 \rm keV}$ (s)             & $2.6\pm0.2$       & $1.17\pm0.04$ & 
 4.7$\pm$0.6 & 1.8$\pm$0.3 & $5.9\pm0.3$& $36.5\pm0.2$\\
\multicolumn{6}{l}{\bf black-body fit}\\
$t_{\rm fit}$ (s)    & 9.3    & 6.0  & 25.1 & 4.0 & -          & 86.4\\
$kT_{\rm bb}$ (kev)  & $2.5\pm0.2$   & $1.1\pm0.14$ & 2.3$\pm$0.2 & 
2.5$\pm$0.3  & 1.6$\pm$0.1 & 2.26$\pm$0.05\\ 
$R$ (km) at $d$      & $4.5\pm0.5$   & $24.^{+5}_{-8}$& 4.7$\pm$0.9 & 
4.7$\pm$0.5 & 5.5$-$11.9 & 6.3$\pm$0.2\\
$F_{2-10 \rm keV}$ ($10^{-8}$ $~\ergcms$) & $1.23\pm0.10$  & 
$1.02\pm0.03$ & 1.08$\pm$0.40& 1.00$\pm$0.42 & $0.4\pm0.1^d$ & 4.05$\pm$0.27\\
$F_{2-28 \rm keV}$ ($10^{-8}$ $~\ergcms$)& $2.17\pm0.07$  & 
$1.07\pm0.05$ & 1.65$\pm$0.73 & 1.66$\pm$0.89 & $0.5\pm0.1^d$ & 6.1$\pm$0.5\\
$F_{\rm peak}$ ($10^{-8}$ $~\ergcms$) & $4.3\pm0.2$ & $1.9\pm0.1$ & 
4.3$\pm$1.6 & 3.3$\pm$1.5 & $3.0\pm1.0^d$ & $9.4\pm1.7$\\
$E_{\rm b}$  ($10^{-7}$ $~\ergcm$) & $\simeq$2.6  &  $\simeq$0.86 &  
$\simeq$4.3 & $\simeq$0.67 & -  &  $\simeq$52 \\
$\chi^2_\nu$ (d.o.f.) & 1.0 (26) & 0.6 (26) & 0.7(26) & 
1.0(26) & 0.9 (270)   & 1.6 (26)\\
\multicolumn{6}{l}{\bf bremsstrahlung  fit}\\
$kT_{\rm brems}$ (keV)& $56^{+47}_{-34}$& $5.9\pm1.7$ 
&  44.$^{+146}_{-22}$ & 50.$^{+149}_{-25}$ & -           &  64$\pm$14\\
$\chi^2_\nu$ (d.o.f.) & 1.4 (26)    & 0.5 (26) & 1.1(26) & 
1.1(26) & -           & 9.7 (26)\\
\multicolumn{5}{l}{\bf power law fit}\\
$\Gamma$              & $1.4\pm0.1$ & $2.3 \pm 0.21$ & 1.4$\pm$0.2 & 
1.4$\pm$0.2 & -       & 1.31$\pm$0.04\\
$\chi^2_\nu$ (d.o.f.) & 1.5 (26)          & 0.5 (26) & 1.1(26) & 
1.1(26) & -           & 10 (26)\\
\hline
\end{tabular}
$^a$\,Interpolated from Dickey \&\ Lockman (1990); $^b$From NFI, see 
Sect. 3.2.1; $^c$\,From Christian \& Swank (1997); $^d$Values for b8.
\end{table*}

\section{Observations and data analysis}

Our observations were obtained from mid 1996 to the end of 2001 with the
Wide Field Cameras (Jager et al. 1997) on board of the BeppoSAX
satellite (Boella et al. 1997).  The Wide Field Cameras are two
identical coded mask aperture cameras with a $40^\circ \times
40^\circ$ (full width to zero response) field of view and $\sim 5'$
angular resolution. The source location accuracy is between $0\farcm7$
and $5'$, and the passband is 2 to 28 keV. Data are collected with a
time resolution of 0.5 ms.  The sensitivity depends on the off-axis
angle, but is on average a few mCrab in a $10^5$ s exposure.

The large sky coverage together with the good angular and time
resolution make the Wide Field Cameras an excellent experiment to
detect fast transient X-ray phenomena at unexpected sky positions and
simultaneously study the behavior of a large fraction of the low mass
X-ray binary population in our Galaxy.

Roughly 90\%\ of all observations of the Wide Field Cameras are
carried out in the so-called 'secondary mode', in which the pointing
of the satellite is set for the Narrow Field Instruments. The pointing
of the Wide Field Cameras is arbitrary, except for solar angle
constraints and the constraint that the two Wide Field Cameras point
in opposite directions, perpendicular to the direction of the Narrow
Field Instruments.  The remaining 10\%\ of the Wide Field Camera
observations are 'primary mode' observations, in which one of the two
cameras is pointed at the Galactic Center (and the other, thus, at the
anti-center).

We search for X-ray bursts in the lightcurve of the total detector
(i.e., a superposition of all the sources in the field of view). With
this method we detect a burst if the fast fluctuations in the overall
count-rate are small ($<10\%$), and if the burst lasts 10 to 100
seconds and reaches a peak count-rate of at least a few times
$10^{-8}$ $\ergcms$. If a burst-like event is observed, a sky image is
reconstructed by cross-correlating the detector image with the coded
mask (Jager et al. 1997).  During this reconstruction the background
is subtracted automatically.  By comparing the sky image with a
catalogue of X-ray sources the burst event can be attributed to a
known, or previously unknown, X-ray burster. In this way we
discovered bursts from SAX\,J1324.5$-$6313, \SAXII, SAX\,J1818.7+1424,
\SAXI\ and 2S\,0918$-$549. The burst positions of the newly discovered
sources are then used to search for persistent emission and other,
possibly somewhat fainter, X-ray bursts in all Wide Field Cameras
observations of these sources.

During primary mode, lightcurves for individual sources are created
and are searched for bursts. In these data we
discovered bursts from 2S\,1711$-$339.

The SAX positions for new X-ray sources are also used to generate
studies with the RXTE All-Sky Monitor (ASM; Levine et al. 1997),
which has smaller cameras than the WFC but typically provides many
measurements per day for every X-ray source. Retrospective ASM light
curves are obtained by re-fitting the coded mask data as a
superposition of the mask shadows for all of the sources in a
particular camera exposure (90 s), including the new target of 
interest.  This reprocessing effort yields light curves that track the
behavior of X-ray sources with substantially greater sensitivity
compared to the threshold for generic identifications of new
transients at random sky positions.

\begin{figure*}[t]
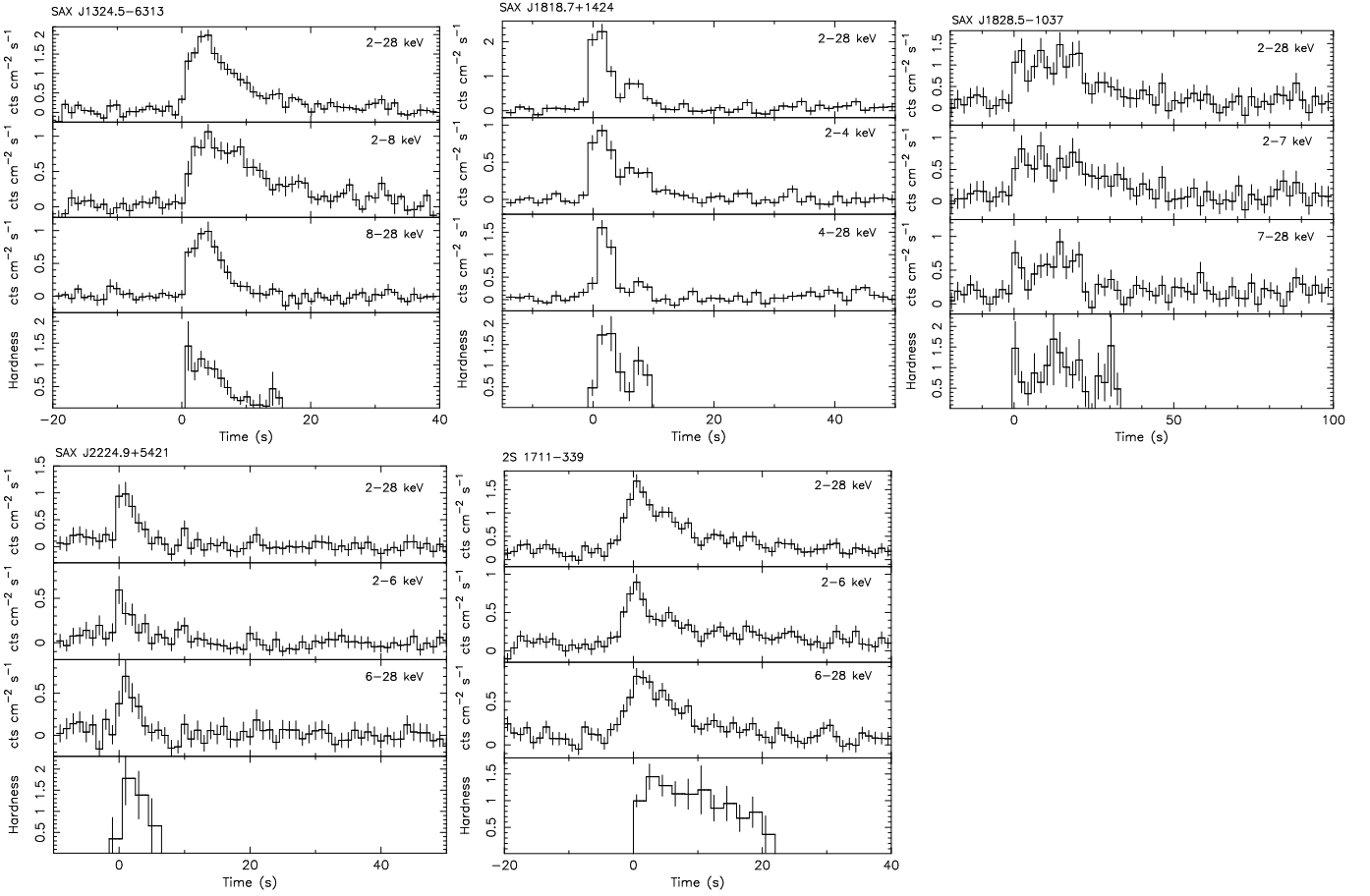

\parbox[b]{6.0cm}{\psfig{figure=H3274.1324.ps,width=6.0cm,angle=-90,clip=t}}
\parbox[b]{6.0cm}{\psfig{figure=H3274.1818.ps,width=6.0cm,angle=-90,clip=t}}
\parbox[b]{6.0cm}{\psfig{figure=H3274.1828.ps,width=6.0cm,angle=-90,clip=t}}
\parbox[b]{6.0cm}{\psfig{figure=H3274.2224.ps,width=6.0cm,angle=-90,clip=t}}
\vspace*{0.5cm}
\parbox[b]{6.0cm}{\psfig{figure=H3274.1711.ps,width=6.0cm,angle=-90,clip=t}}
\parbox[b]{6.0cm}{}
\caption{Lightcurves of bursts from SAX\,J1324.5$-$6313,
SAX\,J1818.7+1424, \SAXII, SAX\,J2224.9$+$5421, and 2S\,1711$-$339, 
respectively. For 2S\,1711$-$339 the profile is the average of 
bursts b1-b10. Three different energy passbands are shown; the 
total Wide Field Cameras passband (2-28 keV), a low energy passband, 
and a high energy passband. The high and low energy passbands 
are chosen in such a way that the count rate was roughly comparable, 
giving the same statistical quality. Also, the hardness, i.e. the
ratio of the high to the low energy count rate, is shown.  For
each source the binsizes are 1 s, 1.5 s, 2 s, 1 s, and 1 s, respectively; 
except for the hardness of SAX\,J2224.9$+$5421 and 2S\,1711$-$339, 
which have binsizes of 2 s. \label{threeS}}
\end{figure*}

\section{Results}

\subsection{New sources}

A single burst was detected at four locations where
no X-ray source is known from previous observations, including the
ROSAT All Sky Survey (Voges et al.\ 1999). We conclude that we have
detected four new sources and designated them with SAX
names in Table\,\ref{results}.

All new sources are detected only during the burst: the
Wide Field Cameras show no persistent emission or other burst
between August 1996 and December 2001 in any of the four cases.  To
compute the upper limits listed in Table\,\ref{results}, we employ
absorption columns interpolated from HI maps by Dickey \& Lockman
(1990), and a power law spectrum with photon index $\Gamma$=1.0. (This
value is typical for burst sources at low luminosities; see for
example Sect. 3.3 below.)  For a photon index of $\Gamma$=2.0 the flux
values would be 70\%\ lower.

The All-sky Monitor onboard RXTE satellite (RXTE/ASM) scanned the
positions of the four new sources. During the course of our analysis
the lightcurves of these sources became available.  These
lightcurves rather uniformly cover a five year period with, most
of the time, several 90 s observations per day and actual exposure
times of $1.7\times10^6$ s and $1.8\times10^6$ s, respectively. None
of the four lightcurves show a clear detection during this period at an
upper-limit of $\simeq2\times10^{-10}$ $\ergcms$ (2-10 keV) in 7 days of
observation.

The lightcurves of the bursts in various passbands are shown in
Fig.\,\ref{threeS}. They have a fast-rise ($\sim$1$-$3\,s)
and exponential-like decay (for \SAXII\ this is evident mainly at
the lower energies). The decay is faster at higher energies,
indicative of spectral softening; decay times are listed in
Table\,\ref{results} together with the spectral fits for the
bursts. 

The burst profile of \SAXII, as shown in Fig.\,\ref{threeS}, looks
rather strange even considering the low statistical quality. But it is
precedented by profiles from at least one well established burster
(4U\,1636$-$536; van Paradijs et~al. 1986). Thus, it would be
interesting to observe more bursts from \SAXII\ with better
sensitivity.

To obtain a spectrum we integrate the spectrum over the burst
duration; the exact integration times are given in
Table\,\ref{results}. We fit the spectra with a black-body, power law,
and bremsstrahlung model. The black-body gives the best
description for three of the four bursts and is acceptable for
SAX\,J1818.7+1424, as expected for type\,I bursts.

If they are indeed type\,I bursts, their peak flux must be less
than or equal to the Eddington limit of $2\times10^{38}\,\ergs$ (for
canonical neutron star values), and we can use the observed peak flux
to obtain an upper limit to the distance (e.g.\ Lewin et al.\ 1993);
these upper limits are also listed in Table\,\ref{results}. We note
that due to systematic uncertainties the errors on the distance are 
$\sim30\%$ (e.g. Kuulkers et al. 2002).

To prove that the above-discussed bursts are genuine type\,I
X-ray bursts it must be shown that a black body gives the only
acceptable description of their spectrum. From Table\,\ref{results} we
see that this is the case only for \SAXIII; for the other three bursts
bremsstrahlung or power-law spectra are still acceptable. That is why
we consider the two alternative explanations for these three bursts,
i.e. that they are stellar X-ray flares or X-ray flashes.

Stellar X-ray flares are generally much longer ($\sim$hours) and have
much lower peak fluxes than the bursts that we have observed
(e.g. Greiner et~al. 1994, Haisch \&\ Strong 1991). We therefore
consider it unlikely that any of the three bursts is a stellar flare.

X-ray flashes are related to gamma-ray bursts in the sense that they
look like the prompt X-ray counterparts to Gamma ray bursts but lack
the $\gamma$-ray emission: they have similar time scales, have the same
variety of time profiles (including fast-rise exponential-decay shapes),
and have spectra that are best described by a power law rather than
black body radiation (Heise et al. 2001). X-ray flashes are
non-repetitive, and if they are related to gamma-ray bursts no
detectable X-ray emission is expected before or lang after the
flash. The Wide Field Cameras have observed about 25 of these
(in 't Zand et al., in prep.). 

\SAXII\ was previously observed during a ROSAT observation (see
Sect. 3.1.1), leaving only doubts on the nature of SAX\,J1818.7$+$1424
and \SAXI.  If these two bursts were X-ray flashes with fast-rise
exponential-decay time profiles, they would be the shortest two of
all, and they would be the only ones for which the black body model
for the spectrum can not be unambiguously ruled out. Together with the
fact that the bursts were at positions near the Galactic plane where
type\,I bursters are likely to occur we think that the most probable
explanation for all four bursts is therefore that they are type\,I
X-ray bursts.

\subsubsection{Other observations} 

HD\,168344, a $V$$=$7.6 magnitude K2-type star is within the
error-radius at a distance of $2\farcm1$ from the centroid of
SAX\,J1818.7+1424. There are 18 stars of magnitude 8 or brighter
within a 4$^\circ\times4^\circ$ field around SAX\,J1818.7+1424 (ESA
1997). This makes the chance probability of having an 8th magnitude or
brighter star within the error radius 0.8\%. This probability is so
small that we consider the possibility that the event was an
X-ray flare from HD\,168344.

According to the Tycho Catalogue (ESA 1997) it is a K2 star with
$V$$=$$7.59$ and $B$$-$$V$$=$$1.047$, a parallax of
$0\farcs0068$$\pm$$0\farcs0055$ and a proper motion of about
0.029$''$/yr.  A main sequence K2 star with this apparent magnitude
would have a distance of only $\sim16$\,pc, incompatible with the
small observed parallax.  In contrast, a K2\,III star of the observed
magnitude would be at a distance of 230-270\,pc, compatible with the
observed parallax, and its velocity perpendicular to the line of sight
would be comparable with the observed radial velocity.

The too short burst time scale of 10\,s and the high peak
luminosity of $1.2\times10^{35}\,\ergs$ at a distance of 230\,pc, both
exclude that the burst observed by us was a stellar flare on
HD\,168344. (No stellar flare this short and bright has ever been
observed to our knowledge; e.g. Haisch \& Strong 1991.)  If a neutron
star were a companion of HD\,168344, the peak flux of its burst would
be $\simeq 6\times10^{-4}$ of the Eddington flux. This also is
unlikely We conclude that HD\,168344 is not the optical counterpart of
SAX\,J1818.7+1424.

Whereas none of the four burst positions coincides with a
catalogued ROSAT source, we find that \SAXII\ is in the field of view
of a 9.4 ks ROSAT/PSPC pointed observation obtained on April 4
1993. We have analyzed this observation and detect seven sources (in
channels 50-240), one of which is in the Wide Field Cameras error-circle.
With a radius for the ROSAT field of view of $\sim50'$ the chance probability that
one of seven sources falls in the \SAXII\ error-circle of $2\farcm8$
is about 2\%. Thus the ROSAT source is probably the counterpart of
\SAXII, confirming that \SAXII\ was not an X-ray flasher. The position
of the ROSAT source is RA=$18^{\rm h} 28^{\rm m}25.7^{\rm s}$,
Dec=$-10^\circ 37' 51''$ (J2000) with an error of $39''$ (1$\sigma$).
The source is not detected in channels 11-50, as expected for a highly
absorbed source. The countrate of 0.011$\pm$0.002 cts s$^{-1}$
corresponds to an unabsorbed flux between 0.5-2.5 keV of
$1.9\times10^{-12}$ $\ergcms$ for a power law with photon index
$\Gamma=1$. At a distance of 6.2 kpc this corresponds to a luminosity
of $8.7\times10^{33}$ $\ergs$.

\begin{figure}[t] 
\psfig{figure=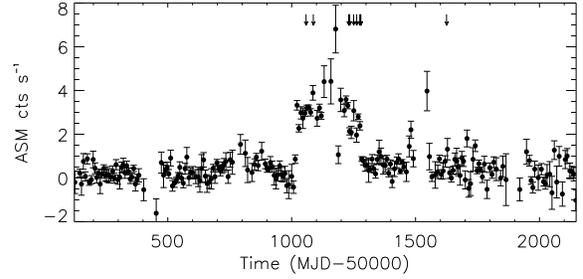,width=8.5cm}
\caption{Long-term lightcurve of 2S\,1711$-$339 obtained with the RXTE
All-sky monitor, which shows an outburst between July 1998 and
May 1999. Each point represents a one-week average. The vertical
arrows indicate the times at which bursts were detected with the
BeppoSAX Wide Field Cameras. \label{asm}}
\end{figure}

\subsection{2S\,1711$-$339}

\begin{figure}[t] 
\psfig{figure=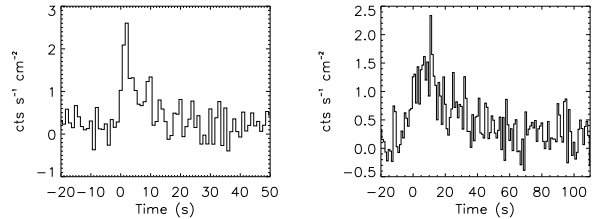,width=8.5cm}
\caption{Bursts b5 (left panel) and b11 (right panel) of 2S\,1711$-$339
in the 2-28 keV passband. The first 10 bursts show similar shapes 
and are represented by b5. The persistent emission 
was detected during the observations before/after all these bursts. 
Before and after b11 the persistent emission was below the detection 
limit of the Wide Field Cameras.}
 \label{burst1711}
\end{figure}

 2S\,1711$-$339 is an X-ray source that was bright between
July 1998 and May 1999. The Wide Field Cameras covered this
outburst in August-October 1998 and February-April 1999.
Ten short bursts were detected which we designate
b1, b2,...,b10. All these bursts appear to have a similar shape.  In
the left panel of Fig.\,\ref{burst1711} we show the best example (b5)
of these bursts. To fit the average persistent emission before and
after the bursts, we employ the absorption column derived in
Sect. 3.2.1. A cut-off power law with a photon-index of 0.7$\pm$0.5
and a high-energy cut off of 2.8$\pm$0.8 keV provides a satisfactory
fit; the flux of the persistent emission is given in
Table\,\ref{results}.

During a 40 ks observation on March 22 2000 an eleventh burst
(b11) was detected while the persistent emission level before and
after the burst was below the detection limit. Assuming that the
spectral parameters of the persistent flux observed with bursts b1-b10
still apply, we obtain a $3\sigma$ upper-limit on the persistent flux
of $7.0\times 10^{-11}$ $\ergcms$ (2-28 keV); this is a factor of ten
lower than during bursts b1-b10. This burst, which is shown in the
right panel of Fig.\,\ref{burst1711}, has a different shape than the
previous 10 bursts. Both the rise and decay time for burst
b11 are longer. In Table\,\ref{1711} we summarize the characteristics
for the 11 individual bursts. In Fig.\,\ref{asm} we show the RXTE/ASM
long-term lightcurve of 2S\,1711$-$339 which shows that the first 10
bursts occurred during an outburst. 

It is clear from Table\,\ref{1711} that none of the exponential fits 
to the decay is formally acceptable. Also, for none of the 
individual bursts cooling can be unambiguously shown. Therefore,
we decided to combine the light curves of burst b1 to b10. The
last burst, b11, is excluded because of its deviating shape. To 
combine the bursts we created lightcurves with a time resolution 
of 1 s. We took the highest bin as the peak for each burst and
the corresponding time as t=0 s. Then the bursts were combined, 
by averaging the count rates and determining the statistical error 
in the mean. Fig.\,\ref{threeS} shows the resulting profile. 

\begin{table}[t]
\caption{Characteristics of the individual bursts of 2S\,1711$-$339.
For each burst we give the time of occurrence, the peak count rate
in the WFC, and from a fit of an exponential to the decay,
the e-folding decay time and the reduced $\chi^2$.}
\label{1711}
\begin{tabular}{lcccc}
\hline
burst & Start time & CR$_{\rm peak}$ & $\tau_{2-28}$ 
& $\chi^2_\nu$\\
    &  (MJD)       &  (cm$^{-2}$ s$^{-1}$) & (s)      & 47 d.o.f.\\  
\hline
b1  & 51058.024301 & 1.15$\pm$0.05 & 6.9$\pm$0.4 & 1.3\\  
b2  & 51087.268967 & 1.73$\pm$0.04 & 7.0$\pm$0.3 & 1.5\\
b3  & 51229.857483 & 1.72$\pm$0.04 & 6.3$\pm$0.2 & 1.3\\
b4  & 51232.083607 & 1.69$\pm$0.04 & 3.6$\pm$0.1 & 1.4\\
b5  & 51234.270915 & 1.63$\pm$0.06 & 5.2$\pm$0.2 & 1.6\\
b6  & 51249.958462 & 0.94$\pm$0.03 &14.5$\pm$0.8 & 1.4\\
b7  & 51262.473439 & 1.29$\pm$0.03 &11.5$\pm$0.4 & 1.4\\
b8  & 51274.775982 & 1.88$\pm$0.04 & 6.9$\pm$0.2 & 1.5\\
b9  & 51278.327360 & 1.25$\pm$0.07 &10.5$\pm$0.9 & 1.5\\
b10 & 51278.992006 & \multicolumn{2}{l}{could not be constrained}\\
b11 & 51625.046991 & 1.18$\pm$0.04 & 15.0$\pm$0.7 & 1.3\\
\hline
\end{tabular}
\end{table}

A black-body spectrum was fit to the burst spectrum of all 11 burst. We
assumed a fixed absorption column, $\nh=1.5\times10^{22}$ $\cmsq$ (see
Sec. 3.2.1). We fitted the 11 burst spectra simultaneously; the
black-body temperature was forced to be the same for all bursts,
whereas the black-body radius was allowed to vary. In
Table\,\ref{results} we have summarized the results for fits to the
spectrum and to the exponential decays in different passbands.
Instead of giving the radius of each individual burst we indicate the
range of radii in Table\,\ref{results}. Varying the temperature and
forcing the radius to be the same for all bursts simultaneously gives
a range of temperatures between 1.2-2.0 keV and a radius of
7.8$\pm$0.9 km (at 7.5 kpc). Burst b8 has the highest peak flux. The
temperatures and the radii found are typical values for Type\,I
bursts. We conclude that the 11 events are Type\,I bursts.

From the observed peak flux during the burst and the constraint that
this must be less than the Eddington limit, we derive from the
brightest burst b8 an upper limit to the distance of 2S\,1711$-$339
of 7.5\,kpc.

\subsubsection{Other observations of 2S\,1711$-$339}

A Chandra/ACIS-S observation of 2S\,1711$-$339 was performed on June 9
2000 for a total of 949 s. A single relatively bright source with a
countrate of 0.46$\pm$0.02 cts s$^{-1}$ is detected at a position
RA=$17^{\rm h}14^{\rm m}19.8^{\rm s}$, Dec=$-34^\circ 02' 47''$
(J2000) with a conservative error-radius of $1''$.

\begin{figure}[t] 
\psfig{figure=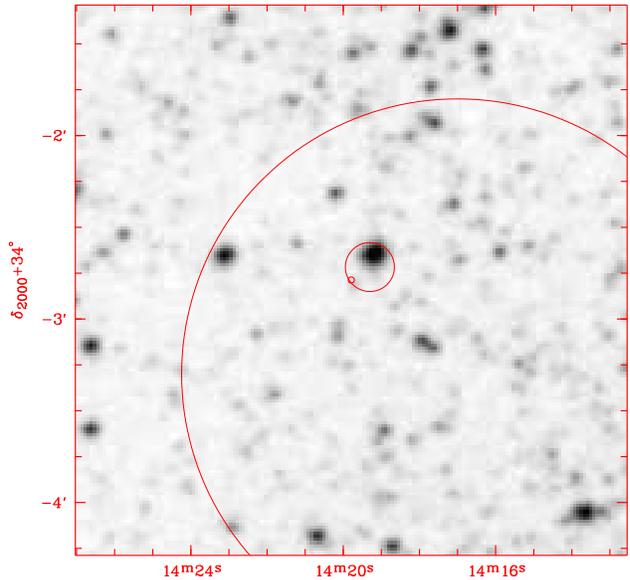,width=\columnwidth,angle=-90,clip=t}
\caption{Error circles (large to small) from the Wide Field Cameras
position (this paper), RXS\,J171419.3$-$340243 (Voges et al.\ 1999),
and Chandra position of 2S\,1711$-$339 (this paper) superposed 
on an image from the Digitized Sky Survey. 
 \label{optic}}
\end{figure}

Our WFC position for 2S\,1711$-$339 (Table\,\ref{results}), the
position of the Ariel\,V source A\,1710$-$34 (Carpenter et al.\ 1977),
and the position of the source RXS\,J171419.3$-$340243 from the ROSAT
All Sky Survey (Voges et al.\ 1999) are all compatible with the
Chandra position of 2S\,1711$-$339 as illustrated in Fig.\,\ref{optic}.
We conclude that they are all the same source. We note that the
nearest source with comparable (ROSAT) brightness in the ROSAT All Sky
Survey is at a distance of $\sim30'$.

The spectrum during a BeppoSAX Narrow Field Instrument (NFI)
observation on February 29, 2000 (MJD 51603) of 2S\,1711$-$339 is best
described by an absorbed power law model with photon index 2.2 and an
absorption column of $1.5\times10^{22}$ $\cmsq$ for the persistent
emission (Migliari, di Salvo, Belloni, in preparation). Use of
this spectrum for the persistent flux measured with the Wide Field
Cameras does not significantly change the numbers given in Sect. 3.2
and Table\,\ref{results}. The flux between 2 and 6\,keV in units of
$10^{-11}$ $\ergcms$ for 2S\,1711$-$339 varied from $<40$ in February
1975 to 200 in September 1976 (Carpenter et al.\ 1977), to 5 during
the EXOSAT Galactic Plane Survey (Warwick et al.\ 1988).  Assuming the
spectrum as observed by the NFI we obtain fluxes between 2 and 6\,keV
in units of $10^{-11}$ $\ergcms$ of 2 during the ROSAT All Sky Survey,
2.4 during the NFI observations, and 0.3 during the Chandra/ACIS-S
observations. The large range of fluxes shows that 2S\,1711$-$339 is
a genuine transient.

\subsection{2S\,0918-549}

A single bright burst was detected from the X-ray source 2S\,0918-549
on June 6.049, 1999. The burst has a fast 
rise, a flat top and an exponential decay. 
In Table\,\ref{results} we summarize exponential-decay fits in
different passbands and
the results of spectral modeling.
We fix the interstellar column at the value found from Einstein
X-ray data (Christian \&\ Swank 1997).

\begin{figure}[t] 
\psfig{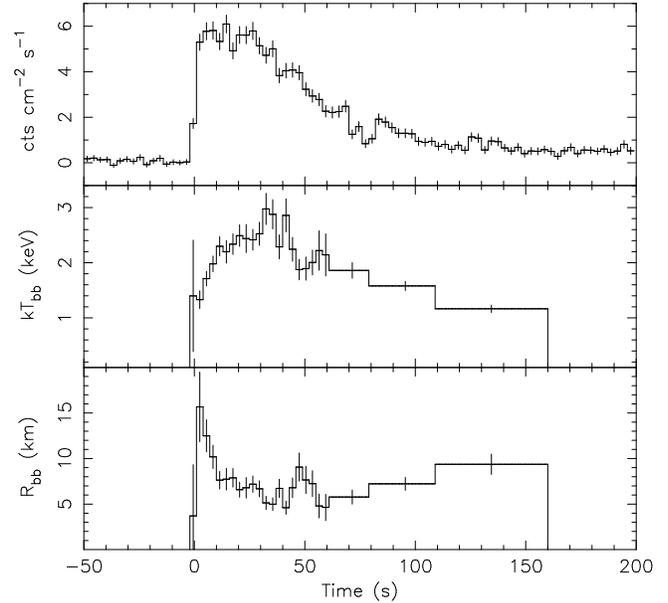}
\caption{The lightcurve and results of the time-resolved spectroscopy
of the burst from 2S\,0918-549. Each bin is 3 seconds, apart from the
last three bins which are 17 s, 30 s and 50 s respectively. The top 
panel gives the lightcurve, the middle panel the black-body temperature 
during the burst and the bottom panel the black-body radius for a 
source at 4.2 kpc. For the spectral fits a fixed column density of 
$\nh=0.5\times10^{22}$ $\cmsq$ is assumed.}
 \label{0918}
\end{figure}

The persistent flux is satisfactorily described by a cut-off power law
spectrum with a photon index of 0.9$\pm$0.6 and a high-energy cut-off
of 5.2$\pm$3.8 keV, or a bremsstrahlung spectrum with a
temperature of 8.9$\pm$1.8 keV (the flux is only marginally different
between the two models).  The black-body model gives the best fit to
the burst spectrum.  We also performed time-resolved spectral fits for
the burst.  The values of $kT_{\rm bb}$ and R$_{\rm bb}$ are plotted
as a function of time in Fig.\,\ref{0918}. We notice that the burst
shows the characteristics of a radius-expansion burst (an increase in
the black-body radius, and a drop in the black-body temperature while
the flux stays constant). A distance of 4.2 kpc (with an uncertainty
of 30\%; see e.g. Kuulkers et~al. 2002) and a persistent luminosity of
$6.8\times10^{35}$ $\ergs$ are implied.

\subsubsection{Other observations}

It turns out that 2S\,0918$-$549 is in the field of view
of a 4.8~ks ROSAT PSPC observation of HD\,81188, with a countrate
(channels 11-240) of 9.83$\pm$0.05 cts~s$^{-1}$. We have analyzed its spectrum
and tried to model it with the combination of a 2~keV black body and
2~keV thermal bremsstrahlung spectrum as fitted by Christian \&\ Swank
(1997) to Einstein data of this source. This model is not acceptable.
An acceptable fit ($\chi^2_\nu=1.2$, 9 d.o.f.)
is obtained for a  combination of a 0.1~keV black body and
3.0~keV thermal bremsstrahlung spectrum, absorbed by
$N_H$$=$$5.0\times 10^{21}$ $\cmsq$.
The ratio of the black-body to the bremsstrahlung flux in the
0.5-2.5 keV band is 1.45. For this fit the 2-10 keV flux is due 
to the bremsstrahlung only, and is $2\times10^{-10}\ergcms$, a 
factor two below the level observed with the Wide Field Cameras.
The flux in the 0.5-20 keV range is  $8\times10^{-10}$ $\ergcms$, a factor 
four higher than the level observed with Einstein.

Jonker et al.\ (2001) were the first to detect a burst of
2S\,0918$-$549. The data analyzed by Jonker et~al.\ do not combine
good time resolution with spectral resolution.  Therefore Jonker
et~al.\ cannot and indeed do not establish that the burst they
observed is at the Eddington-limit. However, they estimate a peak
luminosity of about $8.8\times10^{-8}\ergcms$ (2-20 keV), close to the
peak flux of the burst we observed.  Accordingly, their distance
upper-limit of 4.9 kpc is comparable to our measurement.  We conclude
that the burst observed by Jonker et al.\ must be at, or close to, the
Eddington limit as well; indeed, the lightcurve presented Jonker et
al.\ is highly suggestive of a radius-expansion burst. The upper-limit
$<4.2\times10^{-10}$ $\ergcms$ (2-20 keV) for the persistent flux
before the burst derived by Jonker et al.\ is compatible with (and
close to) the persistent flux that we detect.

The optical counterpart of 2S\,0918$-$549 was identified by
Chevalier \&\ Ilovaisky (1987), as a blue star 
($V$$=$21.0, $B$$-$$V$$=$0.3) in the error
circle of the Einstein position for the X-ray source.
The X-ray to optical flux ratio of $\simeq800$  indicates
that the source is a low-mass X-ray binary.
Chevalier \&\ Ilovaisky assume a distance of 15\,kpc, 
and note that the X-ray and optical luminosities are low, as
compared to those of average low-mass X-ray binaries.
Our distance estimate of 4.2\,kpc leads to lower luminosities 
with $M_{\rm V}$$\simeq$6.9 (for $A_{\rm V}$$\simeq$1.0),
and $L_{\rm x}$$\simeq8.7\times10^{35}\ergs$.

The visual extinction is estimated by assuming that
$(B$$-$$V)_{\rm o}$$=$0 $-$ a common value for low-mass X-ray
binaries $-$ and is roughly compatible with the value estimated
for the X-ray absorption, using the relation between the
latter and the visual extinction according to Predehl
\&\ Schmitt (1995).

\section{Discussion}

 Our observations add at least three and possibly five bursters to
the list of X-ray bursters in our Galaxy and provide a distance
estimate for a recently discovered burster.  In this section we
discuss the implications of our results for the theory of bursts (in
Sect.\ 4.1) and make a comparison between the class of bursters with
low persistent luminosities -- to which we have added four members --
with other low-mass X-ray binaries (in Sect. 4.2).

\subsection{Comparison with burst theory}

We compare the properties of the bursts with theory. Fujimoto et\,al.\
(1987, see also Bildsten 2000) propose three classes of bursts.  At
the lowest accretion rates, $10^{-14}$ M$_\odot{\rm yr}^{-1} \ltap
\dot M \ltap 2\times10^{-10}$ M$_\odot$ yr$^{-1}$, a burst is
triggered by thermally unstable hydrogen burning, and can last between
10$^2$ to 10$^4$ s. At intermediate accretion rates,
$2\times10^{-10}\ltap\dot M\ltap10^{-9}$ M$_\odot$ yr$^{-1}$, a pure
helium burst occurs with a duration of order 10 s.  In the high
accretion regime, $10^{-9}\ltap\dot M\ltap2.6\times 10^{-8} $M$_\odot$
yr$^{-1}$, a burst with a duration of tens of seconds may occur in a
mixed He/H environment. At even higher or lower accretion rates no
bursts are expected to occur.  To consider our observations, we first
converted accretion rates to fractions of the Eddington limit.  The
pure helium bursts occur when the accretion rate is in the range 0.014
- 0.070 of the Eddington accretion rate.  The ratio of (the upper
limit to) the persistent flux and the peak flux during the burst,
where the latter is (a lower limit to) the Eddington flux, provides an
estimate of the fraction of the Eddington limit at which a source is
accreting. This assumes that the emission is isotropic and that the
persistent flux in the range 2-28 keV is close to the bolometric flux.
With the values listed in Table\,\ref{results} we obtain 
$<0.002$ for SAX\,J1324.5$-$6313, 
$<0.009$ for SAX\,J1818.7+1424, 
$<0.004$ for \SAXII, 
$<0.001$ for \SAXI,
$0.02$ for 2S\,1711$-$339 at bursts b1-b10, 
and $<0.002$ at burst b11, 
and $0.004$ for 2S\,0918$-$549.

We thus note that, with the exception of 2S\,1711$-$339 during bursts
b1-b10, all sources are in the low accretion regime, and thus
according to theory should emit bursts lasting longer than about
100\,s.  Source 2S\,1711$-$339 follows this prediction nicely, showing
short ($\simeq$10-20 s) bursts b1-b10 when accretion was in the
intermediate range, and a longer ($\simeq$60 s) burst b11 when the
accretion had dropped to the low regime. Also the burst from
\SAXII\ lasts about 60 s, pointing towards the low accretion regime.
Similarly, the bursts observed by Jonker et al.\ (2001) and by us for
2S\,0918$-$549 are long ($\simeq$150 s), as predicted from the low
accretion rate.

In remarkable contrast, the bursts from SAX\,J1324.5$-$6313,
SAX\,J1818.7+1424 and \SAXI\  are all short ($\simeq$10-20 s),
even though these systems appear to be in the low accretion regime.
Can it be that the true accretion rate is higher than we estimate?
One possibility is that the emission is anisotropic. However, to
our knowledge no indication has been found for anisotropies in other
burst systems. A second possibility would be that most of the
persistent flux is outside the observed 2-28\,keV range. However,
we estimate that more than 50\% of the persistent flux is in this
range. We therefore consider it unlikely that these effects are
sufficient to bring especially SAX\,J1324.5$-$6313 and \SAXI\
to the intermediate accretion regime.

A third possibility is that the accretion is limited to a small area
of the neutron star, e.g.\ a ring connected with the accretion disk
(Popham \& Sunyaev 2001).  This enhances the local accretion rate,
which is the parameter determining the properties of the bursts (as
discussed by Bildsten 2000).  This would imply that the accreting
surfaces in SAX\,J1324.5$-$6313 and \SAXI\ are less than about
15\%\ and 8\% of the surface of the neutron star, 
respectively.  This possibility cannot be excluded {\em a priori},
but raises the interesting question why the accreting surface areas
would be so different between bursters -- the rotation period of the
neutron star could affect the area over which the accreted matter
spreads out, for example.

A fourth possibility is that the persistent flux at the time
of the burst is not representative of the time-averaged flux in the
months before the burst. In transients like e.g. Aql~X-1, Cen~X-4,
XTE J1709$-$267 and SAX\,J1750.8$-$2900, X-ray bursts were
detected during the decline of the outburst, at times when the
persistent flux, easily detectable at $L_{\rm x}\gtap10^{36}$ $\ergs$,
was at an accretion rate of ordinary burst sources (Matsuoka et al.\
1980, Koyama et al.\ 1981, Cocchi et al. 1998, Natalucci et al. 1999).
Also the transient SAX~J1808.4$-$3658 showed a $\sim$$100$ s long
burst 30 days after the peak of an outburst, when the persistent flux
had declined below the detection limit of the Wide Field Cameras,
$<10^{36}$ $\ergs$ (in 't Zand et al.\ 2001). However, the
RXTE/ASM lightcurves show no detection of \SAXIII\ and
SAX\,J1818.7+1424, at an upper-limit of $\simeq10^{36}$ $\ergs$ making
a transient outburst very unlikely.

One might propose the possibility that these systems are old and the
companion has only pure helium left. In this case only helium bursts
can occur independent of the accretion rate. However, calculations on
bursts due to pure helium accretion show that at low accretion rates
the burst duration increases to $\sim100$ s (Bildsten 1995).

For pure helium bursts, the energy released during the burst due to
nuclear fusion is about 1\%\ of the accretion energy released when the
same matter accreted onto the neutron star before the burst (see e.g.\
Lewin et al.\ 1993).  From the observed burst fluences and the (upper
limits to) the persistent flux, we can therefore derive (lower limits
to) the interval to the previous burst.  The computed waiting time
of 16\,d is sufficiently long to explain that only one burst was
detected for 2S\,0918$-$549, whose WFC exposure times totalled for all
observations between August 1996 and December 2001 is about
62\,d. For SAX\,J1324.5$-$6313, SAX\,J1818.7+1424, \SAXII\ and
\SAXI\ the total observation times between August 1996 and
December 2001 are about 58\,d, 30\,d, 25\,d and 65\,d,
respectively.  For these sources the waiting times are $>3.7$\,d,
$>0.6$\,d, $>2.6$\,d and $>2.2$\,d. The chance probability
of observing at most one burst for these sources is then 0.07\% or
(much) less. The fact that only one burst was observed for each
system suggests that the persistent emission levels are much lower
than the upper-limits derived.

\subsection{Low persistent emission bursters}

Gotthelf \&\ Kulkarni (1997) discovered a burst 
from a low-luminosity source in the globular cluster M\,28,
with a peak luminosity that is only 0.02\%\ of the
Eddington limit. This low peak flux discriminates it from
the bursters discussed by Cocchi et al. (2001) and in
this paper, that have fluxes close to the 
Eddington limit: if their peak fluxes were as low
as that of the M\,28 source, they would be a local
population near the Sun, which is clearly incompatible with 
their galactic length and latitude distributions.

As discussed by Cocchi et al.\ (2001), a class of bursters with low
persistent emission has emerged in recent years. The four sources
discussed in Sect. 3.1 also appear to be member of this class,
strenghtening its existence.  Whereas most bursters emit their bursts
at persistent luminosities $\gtap10^{36}$ $\ergs$, most of the members
of this new class emit bursts at luminosities below the RXTE/ASM
detection-limit of $\simeq10^{36}$ $\ergs$. How much lower is not clear, 
and we briefly consider three possibilities. One is that the sources are
steady in the range $10^{34-35}$ $\ergs$, as suggested for the
bursters 1RXS\,J171824.2$-$402934 (Kaptein et al.\ 2000) and \SAXII\
(this paper), whose persistent emission levels were detected at this
level with ROSAT a few years before the burst. The second possibility
is that the sources are steady at the level $10^{32-33}$ $\ergs$, the
quiescent level of soft X-ray transients with neutron stars; and the
third possibility is that they are usually at this low level, but emit
their bursts during or soon after faint ($\ltap10^{36}$ $\ergs$)
outbursts, as suggested by the case of 2S\,1711$-$339.  More sensitive
X-ray observations are required to discriminate between these various
possibilities.

\begin{table}
\caption{Overview of the burst sources at low persistent
emission as observed with the Wide Field Cameras.  
\label{nopersis}}
\begin{tabular}{lc@{\hspace{0.12cm}}l@{\hspace{0.12cm}}l@{\hspace{0.12cm}}c@{\hspace{0.12cm}}}
\hline
Name & $l_{II}$ & $b_{II}$ & $F_{\rm peak}/F_{\rm pers}$  & $\tau$ (s)\\
\hline
SAX\,J1324.5-6313        & $306\fdg64$ & $-0\fdg59$ & $>$540   & 6.0   \\
RX\,J171824.2-402934$^a$ & $347\fdg28$ & $-1\fdg65$ & $>$90    & 47.5  \\
GRS\,1741.9-2853$^b$     & $359\fdg96$ & $0\fdg12$  & $>$130   & 8.8   \\
                         &             &            & $>$180   & 11.0  \\
                         &             &            & $>$100   & 16.0  \\ 
SAX\,J1752.4-3138$^c$    & $358\fdg44$ & $-2\fdg64$ & $>$120   & 21.9  \\
SAX\,J1753.5-2349$^d$    & $5\fdg30$   & $1\fdg10$  & $>$180   & 8.9   \\
SAX\,J1806.5-2215$^d$    & $8\fdg15$   & $-0\fdg71$ & $>$200   & 4.0   \\
                         &             &            & $>$210   & 9.0   \\
\SAXII                   & $20\fdg88$  & $+0\fdg18$ & $>$226   & 11.2  \\
SAX\,J1818.7+1424        & $42\fdg32$  & $13\fdg65$ & $>$110   & 4.5   \\
\SAXI                    & $102\fdg56$ & $-2\fdg61$ & $>$940   & 2.6   \\
\hline
\multicolumn{5}{l}{$^a$ Kaptein et al. (2000); $^b$ Cocchi et al. (1999).}\\
\multicolumn{5}{l}{$^c$ Cocchi et al. (2001); $^d$ in 't Zand et al. (1998).}\\
\end{tabular}
\end{table}

\begin{figure}[t]
\psfig{figure=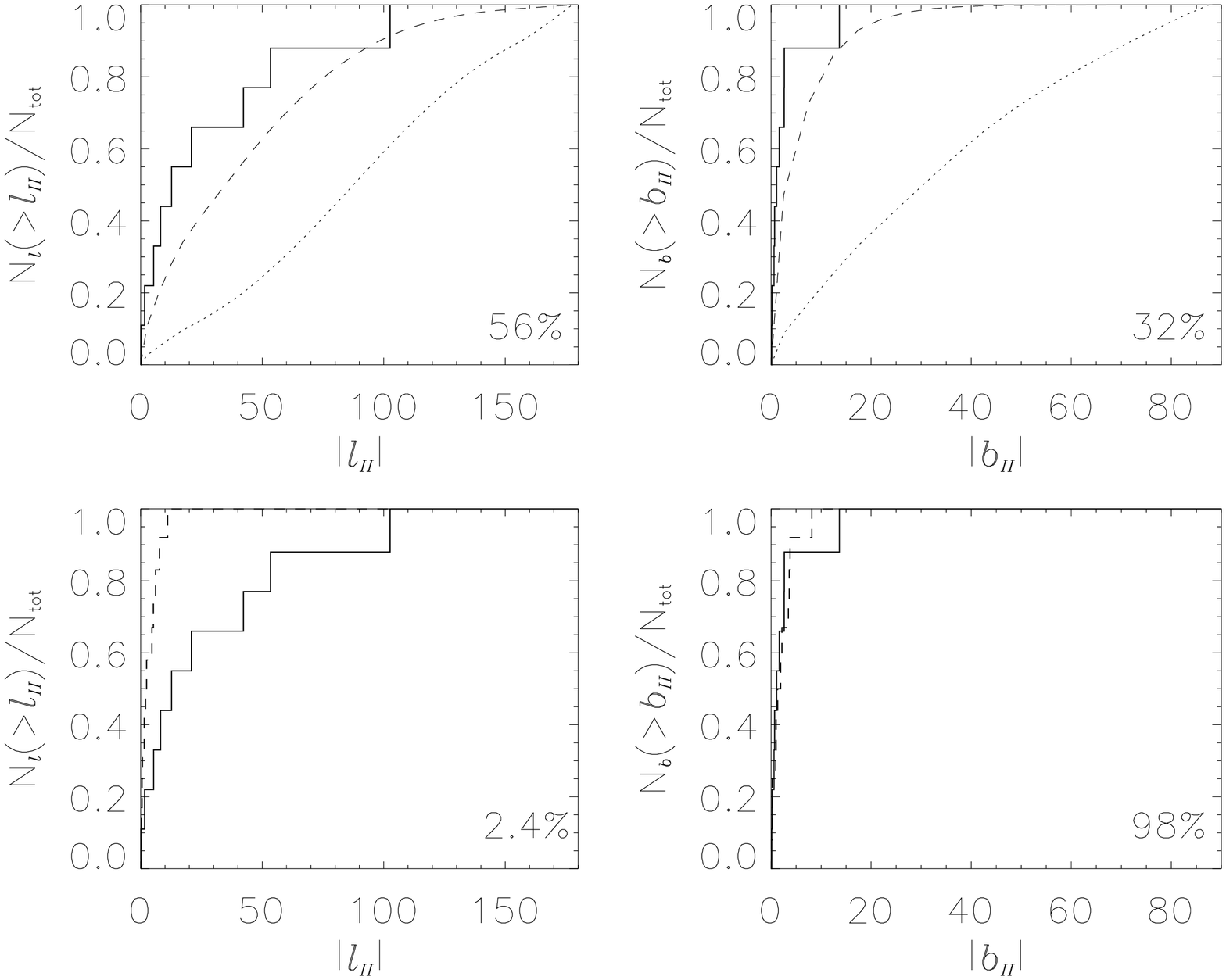,width=9.0cm}
\caption{The top two panels show the cumulative Galactic longitude 
$l_{II}$ and latitude $b_{II}$ distributions (solid lines) of the 
low persistent emission bursters compared to the exponential 
(Galactic) distribution of the low-mass X-ray binaries 
weighted  with observation times (dashed lines). 
The probability according to a two-sided Kolmogorov-Smirnov test
that they have the same distribution is given in the lower right 
corners. For comparison we also show an isotropic distribution 
weighted  with observation times (dotted lines). In the bottom 
two panels the low persistent emission 
bursters (solid line) are compared to the faint transients 
(dashed line). 
\label{kstest}}
\end{figure}

 However, a first test can be made on the basis of the
spatial distributions. In Figure\,\ref{kstest} we compare the
distributions of galactic length and latitude for the
bursters with low persistent luminosity -- listed in Table\,\ref{nopersis}
-- with those of the low-mass X-ray binaries. For the latter we use
exponential distributions in galactic longitude and latitude with
scale angles of $45^{\circ}$ and $8\fdg3$, respectively, as determined
by van Paradijs \&\ White (1995).
Kolmogorov-Smirnov tests indicate that the bursters with low persistent 
luminosity may indeed be drawn from the distribution of low-mass X-ray
binaries. 

A new class of faint transients has been discovered with BeppoSAX:
transients whose outbursts are rather fainter (peaking below $10^{37}$
$\ergs$) and often also shorter (lasting less than a month) than the
outbursts of the ordinary soft X-ray transients which reach the
Eddington limit and may last months (Heise et al. 2000). It is
tempting to assume that the bursters at low persistent emission are
an extension of this class of faint transients.  Remarkably, this new
class of faint transients is more concentrated towards the galactic
center than the ordinary low-mass X-ray binaries (In 't Zand 2001).  A
Kolmogorov-Smirnov test shows that the galactic longitudes of bursters
with low persistent emission cannot be drawn from the longitude
distribution of the faint transients, as illustrated in
Fig.\,\ref{kstest}.

We conclude
that the bursters at low persistent emission are probably not from
the same class as the faint transients, which makes the transient
explanation even more unlikely.

\begin{acknowledgements}
We thank Gerrit Wiersma for discussions on the Wide Field Cameras
exposures. We thank Darragh O'Donoghue for optical observations of 
the bright star near the position of 2S\,1711$-$339. The BeppoSAX
satellite is a joint Italian and Dutch program. We made use of
quick-look results provided by the ASM/RXTE team.
\end{acknowledgements}

\end{document}